\documentclass{elsart3p}

\usepackage{graphicx}
\usepackage{graphics}
\usepackage{amssymb,amsmath}
\usepackage{bm}

\begin{document}

\begin{frontmatter}



\title{Spectrum of Andreev bound states in Josepshon junctions with a ferromagnetic insulator}


\author[address1,address2]{Shiro Kawabata},
\author[address3]{Yukio Tanaka},
\author[address4]{Alexander A.~Golubov},
\author[address5]{Andrey S.~Vasenko},
\author[address7]{Yasuhiro Asano}

\address[address1]{Nanosystem Research Institute (NRI), National Institute of Advanced Industrial Science and Technology (AIST), Tsukuba, Ibaraki 305-8568, Japan}

\address[address2]{CREST, Japan Science and Technology Corporation (JST), Kawaguchi, Saitama 332-0012, Japan}

\address[address3]{Department of Applied Physics, Nagoya University, Nagoya, 464-8603, Japan}

\address[address4]{Faculty of Science and Technology, University of Twente, P.O. Box 217, 7500 AE Enschede, The Netherlands}

\address[address5]{Institut Laue-Langevin, 6 rue Jules Horowitz, BP 156, 38042, Grenoble, France}

\address[address7]{Department of Applied Physics, Hokkaido University, Sapporo, 060-8628, Japan}

\begin{abstract}
Ferromagnetic-insulator (FI) based Josephson junctions are promising candidates for a coherent superconducting quantum bit as well as a classical superconducting logic circuit.
Recently the appearance of an intriguing atomic-scale 0-$\pi$ transition has been theoretically predicted.
In order to uncover the mechanism of this phenomena, we numerically calculate the spectrum of Andreev bound states in a FI barrier by diagonalizing the Bogoliubov-de Gennes equation.
We show that  Andreev spectrum  drastically depends on the parity of the FI-layer number $L$ and accordingly the $\pi$ (0) state is always more stable than the 0 ($\pi$) state if $L$ is odd (even). 
\end{abstract}

\end{frontmatter}

\newpage
\section{Introduction}

The peculiarity of the proximity effect in superconductor/ferromagnetic-metal (S/FM) bilayers is the damped oscillation of the pair amplitude inside a FM~\cite{rf:Bulaevskii,rf:Buzdin2}.
This anomalous proximity effect leads to the $\pi$ Josephson S/FM/S junction~\cite{rf:Golubov,rf:Buzdin1}  which has the opposite sign to the superconducting order parameter in two S electrodes in the ground state. 
Experimentally $\pi$-junction was firstly observed by Ryazanov~\cite{rf:Ryanzanov} and Kontos~\cite{rf:Kontos} and since then a lot of progress has been made in the physics of $\pi$-junctions and now they are proving to be promising elements of superconducting classical and quantum circuits~\cite{rf:Ioffe,rf:Petkovic,rf:Krasnov,rf:Ustinov}.

On the other hands, recently a possibility of $\pi$ junction formation in a Josephson junction with a $ferromagnetic$ $insulator$ (FI) has been theoretically predicted~\cite{rf:Tanaka,rf:Cuevas,rf:Zhao,rf:Kawabata1,rf:Kawabata2,rf:Kawabata3,rf:Kawabata4,rf:Kawabata5,rf:Kawabata5-2,rf:Kawabata6}.
The $\pi$ junction using such an insulating barrier is very promising for future qubit~\cite{rf:Kawabata8,rf:Kawabata9,rf:Kawabata10,rf:Kawabata11} and microwave~\cite{rf:Hikino} applications because of the low decoherence nature~\cite{rf:Zaikin,rf:Kato}.
More importantly, it has been shown that the ground state of S/FI/S junction alternates between 0- and $\pi$-states when thickness of FI
is increasing by a single atomic layer~\cite{rf:Kawabata3,rf:Kawabata5}.
In this paper in order to understand the physical mechanism of the anomalous atomic scale 0-$\pi$ transition, we will calculate the spectrum of the Andreev bound states in such systems.
Based on this calculation, we will show that Andreev spectrum  drastically depends on the parity of the FI layer number $L$ and thence the $\pi$ (0) state is always more stable than the 0 ($\pi$) state if $L$ is odd (even).

In this paper we focused on the one dimensional $s$-wave junction with a FI barrier (Fig. 1(a)).
It should be noted that the qualitatively same result can be obtained for two- or three-dimensional cases.

\section{Model}
\label{}
%
%
%
%
%
\begin{figure}[b]
\begin{center}
\includegraphics[width=8.0cm]{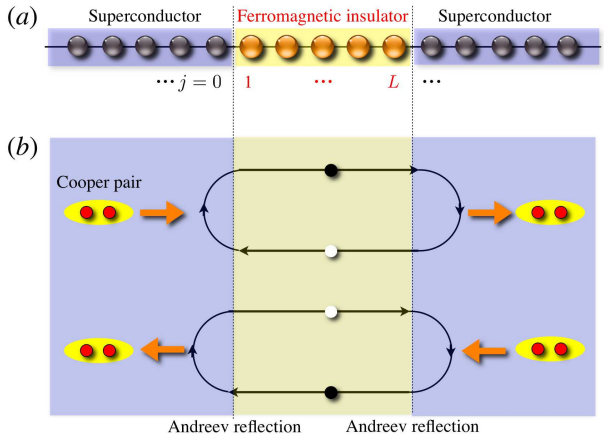}
\end{center}
\caption{(a) The Josephson junction with a ferromagnetic-insulator barrier on the one-dimensional tight-binding lattice and (b) transport  through Andreev bound states. An electron (black circle) is reflected as a hole (white circle) at the interface with the left superconductor and the hole is reflected back as an electron on the right interface. The net result is a Cooper pair transfer through the junction.
}\label{fig1}
\end{figure}

Let us consider a one-dimensional tight-binding lattice of a superconductor/ferromagnetic-insulator /superconductor (S/FI/S) Josephson junction with $L$ being the thickness or the numbers of the FI lattice sites as shown in Fig.~1(a).
The lattice constant is set to be unity.
Electronic states in a $s$-wave superconductor are described by the mean-field BCS Hamiltonian, 
\begin{eqnarray}
H_\mathrm{S}
&=&
-t
\sum_{n,n^\prime \in \mathrm{S}, \sigma} 
c_{n \sigma }^\dagger
c_{n^\prime \sigma }
+
(2 t -\mu_s) 
\sum_{n \in \mathrm{S} , \sigma} 
c_{n \sigma }^\dagger
c_{n \sigma }
\nonumber\\
&+&
\frac{ \Delta}{2} 
\sum_{n \in \mathrm{S}}
\left(
   c_{n \uparrow }^\dagger
   c_{n \downarrow }^\dagger
+ 
   c_{n \uparrow }^\dagger
   c_{n \downarrow }^\dagger
+
 \mathrm{h.c.}
 \right)
.
\end{eqnarray}
Here
  $
  c_{n \sigma }^{\dagger }$ ($c_{n \sigma }^{{}}
$)
 is the creation
(annihilation) operator of an electron at a site $n \in \mathrm{S}$ with spin
 $\sigma
=$ ( $\uparrow $ or $\downarrow $ ) and $\mu_s$ is the chemical potential.
The hopping integral $t$ is considered among nearest neighbor sites and $\Delta$ is the amplitude of $s$-wave pair potential.

The energy dispersion in the Bogoliubov-de Gennes picture and the spin resolved density of states (DOS) for typical FIs are shown schematically in Fig. 2.
Experimental studies as well as a first principle calculations indicate that the band structure of an oxide ferromagnet La${}_4$Ba${}_2$Cu${}_2$O${}_{10}$ (La422)~\cite{rf:Mizuno,rf:LBCO1,rf:LBCO2} and K${}_2$CuF${}_{4}$~\cite{rf:KCF1,rf:KCF2,rf:KCF3} can be described by Fig. 2(b) in which the up- and down-spin bands are located below and above the Fermi energy respectively.
The exchange splitting $V_\mathrm{ex}$ of La422 is numerically estimated to be 0.34 eV.
Since the exchange splitting is large and the bands are originally half-filled, La422 becomes FI with a Curie temperature of 5 K~\cite{rf:Mizuno}. 
Another possible candidates for the FI barrier are spinels~\cite{rf:spinel1,rf:spinel2}, e.g., NiFe${}_{2}$O${}_{4}$, rare-earth monopnictides~\cite{rf:GdN1,rf:GdN2,rf:GdN3,rf:GdN4}, e.g., GdN, and Yttrium iron garnet (Y${}_{3}$Fe${}_{5}$O${}_{12}$)~\cite{rf:YIG1,rf:YIG2}. 

\begin{figure}[b]
\begin{center}
\includegraphics[width=8.0cm]{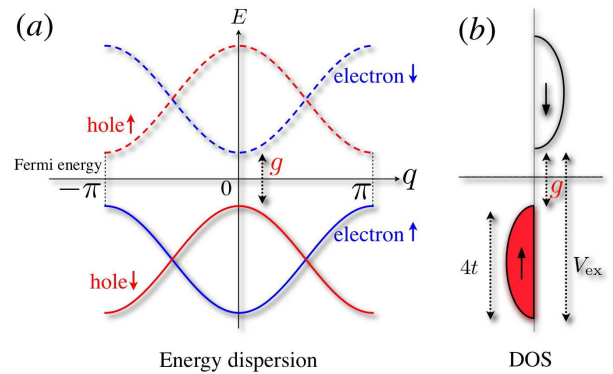}
\end{center}
\caption{(a)The band structure of a ferromagnetic-insulator in the Bogoliubov-de Gennes picture and (b) corresponding spin-resolved density of states (DOS).
The energy dispersion for a hole with spin $\sigma$ is obtained as a mirror image of that for an electron with spin $\sigma$ with respect to Fermi energy.}
\label{fig1}
\end{figure}

The Hamiltonian of a ferromagnetic layer can be described by a single-band tight-binding model~\cite{rf:Kawabata6} as 
\begin{eqnarray}
H_\mathrm{FI}
& =& -t \sum_{n,n^{\prime } \in \mathrm{F},\sigma} 
c_{n\sigma}^\dagger 
c_{n'\sigma}
\nonumber\\
&-&
\sum_{n \in \mathrm{F}} \left( 4 t -\mu + \frac{V_\mathrm{ex}}{2} \right)  
c_{n \uparrow}^\dagger 
c_{n \uparrow}
\nonumber\\
&+&
 \sum_{n \in \mathrm{F}} 
\left( 4 t -\mu + \frac{V_\mathrm{ex}}{2}  \right)  
 c_{n\downarrow}^\dagger 
 c_{n\downarrow}
,
\end{eqnarray}
where 
\begin{eqnarray}
V_\mathrm{ex}=4 t+ g
\end{eqnarray}
is the exchange splitting  ($g$ is the gap between up  and down spin bands) and $\mu$ is the chemical potential (see Fig. 2(a)).
If $V_\mathrm{ex} >  4 t$, this Hamiltonian describes FI as shown in Fig. 2.

\section{Andreev bound states and Josephson current}
\label{}

The Hamiltonian can be diagonalized by the Bogoliubov transformation.
Due to the the Andreev reflection at S/FI interfaces, the Andreev bound state is formed in the FI barrier (see Fig. 1(b)).
Wave functions of the Andreev bound state decay far from the S/FI interface.
In what follows, we focus on the subspace for
spin-$\uparrow$ electron and spin-$\downarrow$ hole.
In superconductors, the wave function of a bound state is given by
\begin{eqnarray}
\Psi_{L}(n)=&\Phi_L \left[
 \left(\begin{array}{c} u \\ v \end{array} \right)
 A e^{-ikn} +
 \left(\begin{array}{c} v \\ u \end{array} \right)B e^{ik^* n}\right] ,
 \\
\Psi_{R} (n)=&\Phi_R \left[
 \left(\begin{array}{c} u \\ v \end{array} \right)
 C e^{ikn} +
 \left(\begin{array}{c} v \\ u \end{array} \right)D e^{-ik^* n}\right] .
 \end{eqnarray}
Here $A, B, C$ and $D$
are amplitudes of the wave function for an outgoing quasiparticle,  $\phi_\nu$ is the phase of a superconductor,
\begin{eqnarray}
\Phi_\nu&=&\mathrm{diag} \left(  e^{i\phi_\nu/2} , e^{-i\phi_\nu/2} \right)
,
\end{eqnarray}
with $\nu=L$ ($R$) indicates an left (right) superconductor, and
\begin{eqnarray}
u&=&\sqrt{\frac{1}{2} \left(  1+ \frac{\Omega}{E} \right)}
\\
v&=&\sqrt{\frac{1}{2} \left(  1- \frac{\Omega}{E} \right)}
,
\end{eqnarray}
with 
\begin{eqnarray}
\Omega= \sqrt{E^2 - \Delta^2}
.
\end{eqnarray}
The energy $E$ is measured from the Fermi energy
and 
\begin{eqnarray}
k=\frac{\pi}{2}
+ i \cosh^{-1} 
\sqrt{1+ \frac{\Delta^2 - E^2}{4 t^2}}
\end{eqnarray}
is the complex wave number.
In a FI, the wave function is given by
\begin{eqnarray}
\Psi_\mathrm{FI}(n)&=&
 \left(\begin{array}{c} f_1 e^{-iq_e n} \\ g_1 e^{-iq_h n}\end{array} \right)
+\left(\begin{array}{c} f_2 e^{iq_e n} \\ g_2 e^{iq_h n}\end{array} \right)
,
\end{eqnarray}
with
\begin{eqnarray}
 q_e&=&\pi + i  \cosh^{-1} \left( 1 + \frac{E}{2 t} + \frac{g}{4 t} \right), \label{wn1}\\
   q_h&=&i  \cosh^{-1} \left( 1 + \frac{E}{2 t} + \frac{g}{4 t} \right), \label{wn2}
\end{eqnarray}
and
$f_1, f_2, g_1$ and $g_2$ are amplitudes of wave function in a FI.

By applying the boundary conditions, 
\begin{eqnarray}
\Psi_L(0)&=&\Psi_\mathrm{FI}(0),
\\
\Psi_L(1)&=&\Psi_\mathrm{FI}(1),
\\
\Psi_R(L)&=&\Psi_\mathrm{FI}(L),
 \\
\Psi_R(L+1)&=&\Psi_\mathrm{FI}(L+1)
,
 \end{eqnarray}
we can obtain a secular equation for amplitudes $A$, $B$, $C$ and $D$. 
From this equation, we can numerically calculate the Andreev levels $\varepsilon_{j}$ as a function of the phase difference $\phi= \phi_L -\phi_R$, where $j=1, \cdots, 4$.

The Josephson current can be  calculated from the Beenakker formula~\cite{rf:Beenakker}, $i.e.,$
\begin{eqnarray}
 I_J (\phi) =\frac{2e }{ \hbar} \sum_{j}  
 \frac{\partial \varepsilon_{j} (\phi)}
{ \partial \phi }
 f  \left[ \varepsilon_{j} (\phi) \right]
 ,
  \end{eqnarray}
where $ f\left( \varepsilon  \right)$ is the Fermi-Dirac distribution function.
In the case of a high barrier limit, the Josephson current phase relation is described by 
\begin{eqnarray}
I_J(\phi) = I_C \sin \phi.
  \end{eqnarray}
Thus we define the Josephson critical current $I_C$ as  
\begin{eqnarray}
I_C = I_J\left( \frac{\pi}{2} \right)
.
\end{eqnarray}
If $I_C$ is negative (positive), then the $\pi$ (0) junction is realized.

\section{Numerical results}
In this section, we show numerical results for the spectrum of Andreev bound states for a conventional S/I/S junction and an S/FI/S junction.
In the calculation, we set $\mu=\mu_s=2t$, and $\Delta=0.01 t$.
\begin{figure}[tb]
\begin{center}
\includegraphics[width=8.0cm]{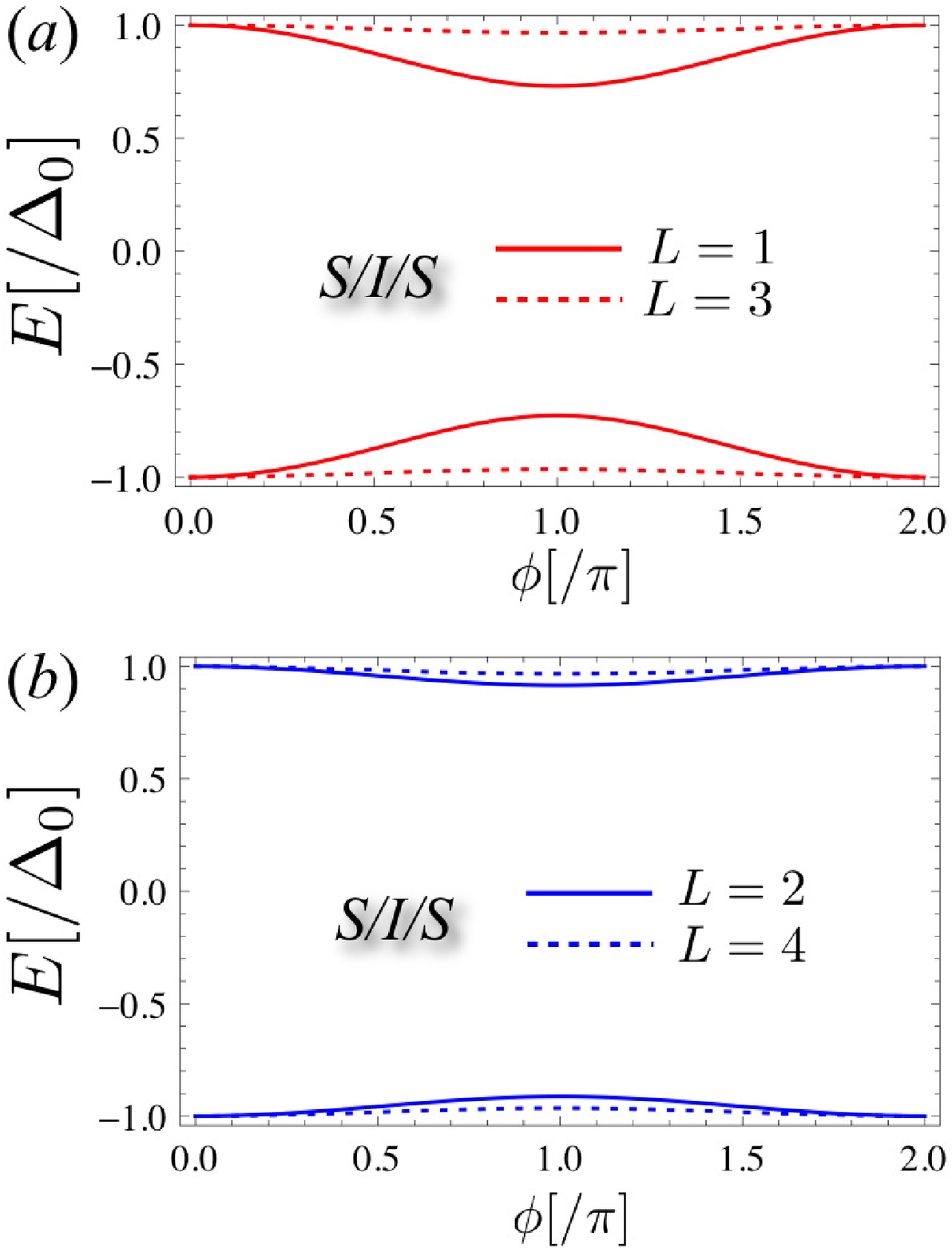}
\end{center}
\caption{The Andreev spectrum in an S/I/S Josephson junction for the case of (a) odd  and (b) even insulating-layer number $L$. 
In the calculation, we set $g=0.25 t$, $\mu=\mu_s=2t$, and $\Delta=0.01 t$.}
\label{fig2}
\end{figure}
\begin{figure}[tb]
\begin{center}
\includegraphics[width=8.0cm]{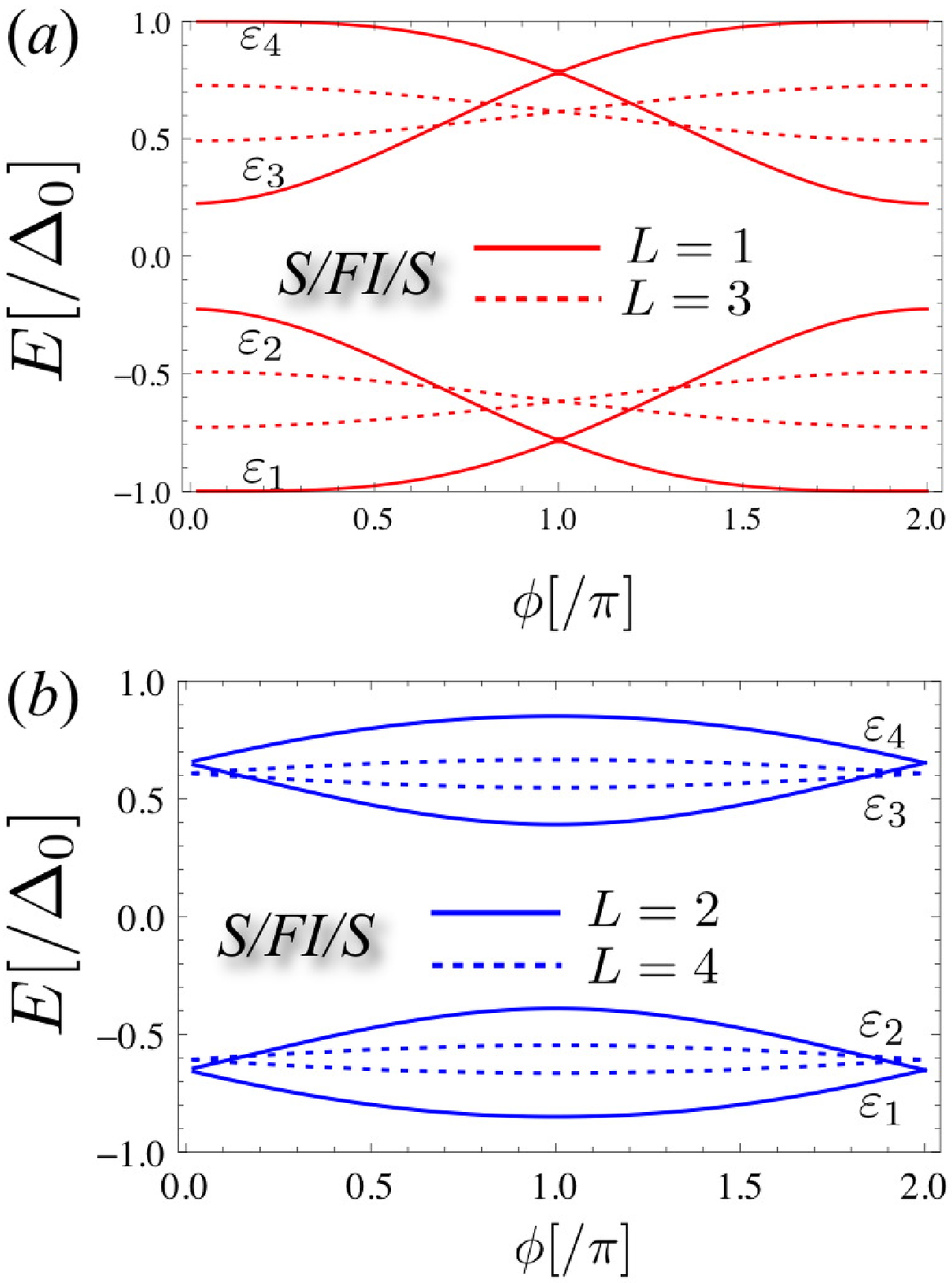}
\end{center}
\caption{The Andreev spectrum in an S/FI/S Josephson junction for the case of (a) odd  and (b) even FI-layer number $L$. 
In the calculation, we set $g=t$, $\mu=\mu_s=2 t$, and $\Delta=0.01 t$.}
\label{fig3}
\end{figure}

Let us firstly consider Andreev bound states in an S/I/S junction.
Fig. 3 shows the Andreev spectrum as a function of the thickness of the insulating barrier $L$.
Due to the spin degeneracy, we have 2 Andreev levels for a given $\phi$ and $L$.
It is evident  that the energy minimum is at $\phi=0$ irrespective of the value of $L$.
So the overall feature of Andreev levels does not depend on $L$. 
On the other hand, Fig. 4 shows the $L$ dependence of the Andreev spectrum for an S/FI/S junction.
The results indicate that the overall feature of the spectrum strongly depends on the parity of $L$ and show that the energy minimum of $\varepsilon_1$ for odd $L$ is at $\phi=0$, whereas for even $L$  at $\phi=\pi$.

\begin{figure}[bt]
\begin{center}
\includegraphics[width=8.0cm]{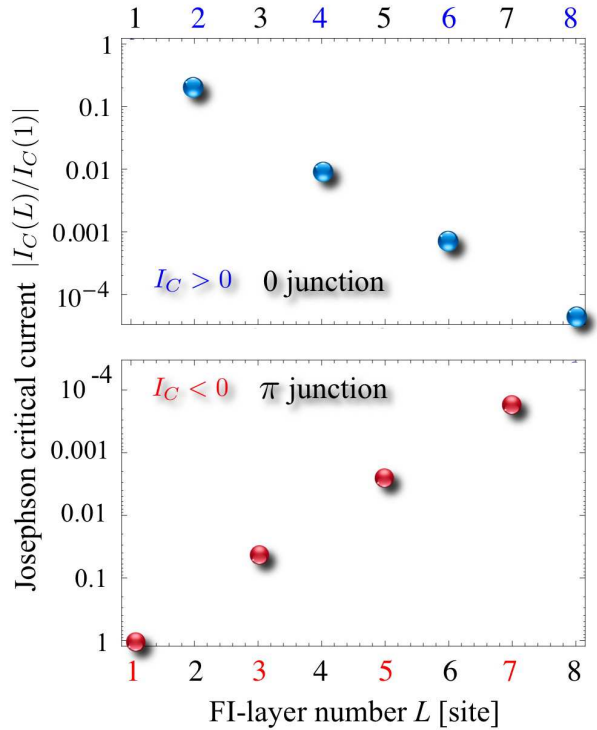}
\end{center}
\caption{Josephson critical current  $I_C$ at $T=0.01 T_c \ll T_c$ as a function of the FI layer thickness $L$ for $g=t$. The red (blue) circles indicate the $\pi$(0) junction. }
\label{fig4}
\end{figure}

In Fig.~5, we show the Josephson critical current for an S/FI/S junction as a function of $L$.
Temperature $T$ is set to be $0.01T_c  \ll T_c$, where $T_c$ is the transition temperature of a superconductor.
The $\pi$(0)-state is always more stable than the 0($\pi$)-state when the thickness of FI is an odd(even) integer.
Based on the Andreev spectrum (Fig. 4) the reason can be explained as follows.
At low temperatures, only the Andreev levels below the Fermi energy i.e.,
$\varepsilon_{1}$ and
$\varepsilon_{2}$, contribute to $I_C$.
In the odd (even) $L$ cases, the $\pi$- (0-) junction is stable because of
\begin{eqnarray}
\left. \frac{\partial \varepsilon_1 (\phi) }{\partial \phi} \right|_{\phi=\pi/2} &>& (<) 0,
\\
\left. \frac{\partial \varepsilon_2 (\phi) }{\partial \phi} \right|_{\phi=\pi/2}&<& (>) 0,
\\
\left| \frac{\partial \varepsilon_2 (\phi) }{\partial \phi} \right|_{\phi=\pi/2}
&>&
\left| \frac{\partial \varepsilon_1 (\phi) }{\partial \phi} \right|_{\phi=\pi/2}
.
  \end{eqnarray}
Above analysis provides an new physical interpretation of the atomic scale 0-$\pi$ transition from the view point of the Andreev spectrum.

\section{Summary}
To summarize, we have theoretically studied the Andreev levels and the Josephson current in S/FI/S junctions by solving the Bogolubov-de Gennes equation in order to understand the physical mechanism of the atomic scale 0-$\pi$ transition. 
A characteristic and important feature for such systems is that the Andreev spectrum  strongly depends on the parity of the thickness of the FI layer $L$.
As a result, the junctions show the atomic scale 0-$\pi$ transition.
Our finding suggests a way of understanding the physical origin of the atomic scale 0-$\pi$ transition in ferromagnetic-insulator based Josephson junctions. 
In this paper, we have only considered the Josephson transport in the low temperature regime, i.e., $T \ll T_c$.
The calculation of $I_C$ in the finite temperature region and the analysis based on the Andreev spectrum (Fig. 4)  are  important future problems.

\section*{Acknowledgements}
 We would like to thank S. Kashiwaya for useful discussion.
This work was  supported by CREST-JST, the "Topological Quantum Phenomena" (No. 22103002) KAKENHI on Innovative Areas and  a Grant-in-Aid for Scientific Research (No. 22710096) from MEXT of Japan.


\begin{thebibliography}{99}
%
%
%
%
\bibitem{rf:Bulaevskii}
L. N. Bulaevskii, V. V. Kuzii, A. A. Sobyanin, 
JETP Lett. {\bf 25} (1977) 291. 
%
%
\bibitem{rf:Buzdin2}
A. I. Buzdin, L. N. Bulaevskii, S. V. Panyukov, 
JETP Lett. {\bf 35} (1982) 179. 
%
%
\bibitem{rf:Golubov}
A. A. Golubov, M. Y. Kupriyanov, E. Il'ichev, 
Rev. Mod. Phys. {\bf 76} (2004) 411.
%
%
\bibitem{rf:Buzdin1}
A. I. Buzdin,
Rev. Mod. Phys. {\bf 77} (2005) 935.
%
%
\bibitem{rf:Ryanzanov}
V. V. Ryazanov, V. A. Oboznov, A. Y. Rusanov, A. V. Veretennikov, A. A. Golubov, J. Aarts,
Phys. Rev. Lett. {\bf 86} (2001) 2427.
%
%
\bibitem{rf:Kontos}
T. Kontos, M. Aprili, J. Lesueur, F. Gen\^et, B. Stephanidis, R. Boursier,
Phys. Rev. Lett. {\bf 89} (2002)137007.
%
%
\bibitem{rf:Ioffe}
L.B. Ioffe, V.B. Geshkenbein, M.V. Feigelman, A.L. Fauchere, G. Blatter,
Nature {\bf 398} (1999) 679.
%
%
\bibitem{rf:Petkovic}
I. Petkovic, M. Aprili, 
Phys. Rev. Lett. {\bf 102} (2009)157003.
%
%
\bibitem{rf:Krasnov}
V. M. Krasnov, T. Golod, T. Bauch, and P. Delsing, 
Phys. Rev. B 76 (2007) 224517.
%
%
\bibitem{rf:Ustinov}
A. K. Feofanov, V. A. Oboznov, V. V. Bolginov, J. Lisenfeld, S. Poletto, V. V. Ryazanov, A. N. Rossolenko, M. Khabipov, D. Balashov, A. B. Zorin, P. N. Dmitriev, V. P. Koshelets, A. V. Ustinov,
Nature Phys. {\bf 6} (2010) 593.
%
%
\bibitem{rf:Tanaka}
Y. Tanaka, S. Kashiwaya,
Physica C {\bf 274} (1997) 357.
%
%
\bibitem{rf:Cuevas}
J. C. Cuevas, M. Fogelstr\"om,
Phys. Rev. B 64 (2001) 104502.
%
%
\bibitem{rf:Zhao}
E. Zhao, T. L\"ofwander, and J. A. Sauls
Phys. Rev. B 70 (2004) 134510.
%
%
\bibitem{rf:Kawabata1}
S. Kawabata, Y. Asano,
Int. J. Mod. Phys. B {\bf 23} (2009) 4329.
%
%
\bibitem{rf:Kawabata2}
S. Kawabata, Y. Asano, Y. Tanaka, S. Kashiwaya,
Physica C {\bf 469} (2009) 1621.
%
%
\bibitem{rf:Kawabata3}
S. Kawabata,  Y. Asano, Y. Tanaka, A. A. Golubov, S. Kashiwaya,
Phys. Rev. Lett.  {\bf 104} (2010) 117002.
%
%
\bibitem{rf:Kawabata4}
S. Kawabata, Y. Asano, Y. Tanaka, S. Kashiwaya, 
Physica E  {\bf 42} (2010) 1010.
%
%
\bibitem{rf:Kawabata5}
S. Kawabata, Y. Asano,
Low Temp. Phys.  {\bf 36} (2010) 915.
%
%
\bibitem{rf:Kawabata5-2}
S. Kawabata, Y. Asano, Y. Tanaka, A. A. Golubov, S. Kashiwaya,
Physica C {\bf 470}  (2010) 1496.
%
%
\bibitem{rf:Kawabata6}
S. Kawabata, Y. Tanaka,  Y. Asano,
Physica E {\bf 43} (2011) 722.
%
%
\bibitem{rf:Kawabata8}
S. Kawabata, S. Kashiwaya, Y. Asano, Y. Tanaka,
Physica C {\bf 437-438} (2006) 136.
%
%
\bibitem{rf:Kawabata9}
S. Kawabata, S. Kashiwaya, Y. Asano, Y. Tanaka, A. A. Golubov,
Phys. Rev. B  {\bf 74} (2006) 180502(R).
%
%
\bibitem{rf:Kawabata10}
S. Kawabata, A. A. Golubov,
Physica E {\bf 40} (2007) 386.
%
%
\bibitem{rf:Kawabata11}
S. Kawabata, Y. Asano, Y. Tanaka, S. Kashiwaya, A. A. Golubov,
Physica C {\bf 468} (2008) 701.
%
%
\bibitem{rf:Hikino}
S. Hikino, M. Mori, S. Takahashi, and S. Maekawa,
J. Phys. Soc. Jpn. {\bf 80} (2011) 074707.

%
\bibitem{rf:Zaikin}
G. Sch\"on, A. D. Zaikin, 
Phys. Reports {\bf 198} (1990) 237.
%
%
\bibitem{rf:Kato}
T. Kato, A. A. Golubov, Y. Nakamura, 
Phys. Rev. B {\bf 76} (2007) 172502.
%
%
\bibitem{rf:Mizuno}
F. Mizuno, H. Masuda, I. Hirabayashi, S. Tanaka, M. Hasegawa, U. Mizutani,
Nature {\bf 345} (1990) 788.
%
%
\bibitem{rf:LBCO1}
V. Eyert, K. H. H\"oc, P. S. Riseborough,
Europhys. Lett. {\bf 31} (1995) 385.
%
%
\bibitem{rf:LBCO2}
W. Ku, H. Rosner, W. E. Pickett, R. T. Scalettar,
Phys. Rev. Lett.  {\bf 89} (2002) 167204.
%
%
\bibitem{rf:KCF1}
I. Yamada, 
J. Phys. Soc. Jpn. {\bf 33} (1972) 979.
%
%
\bibitem{rf:KCF2}
K. Hirakawa, H. Ikeda, 
J. Phys. Soc. Jpn. {\bf 35} (1973) 1328.
%
%
\bibitem{rf:KCF3}
V. Eyert, K. H. H\"oc,
J. Phys.: Condens. Matter {\bf 5} (1993) 2987.
%
%
\bibitem{rf:spinel1}
Z Szotek, W. M. Temmerman, A. Svane, L. Petit, P. Strange, G. M. Stocks, D. K\"odderitzsch, W. Hergert, H. Winter,
J. Phys.: Condens. Matter {\bf 16} (2004) S5587.
%
%
\bibitem{rf:spinel2}
Z. Szotek, W. M. Temmerman, D. K\"odderitzsch, A. Svane, L. Petit, H. Winter,
Phys. Rev. B {\bf 74} (2006) 174431.
%
%
\bibitem{rf:GdN1}
C-G. Duan, R. F. Sabirianov, W. N. Mei, P. A. Dowben, S. S. Jaswal, E. Y. Tsymbal,
J. Phys.: Condens. Matter {\bf 19} (2007) 315220.
%
%
\bibitem{rf:GdN2}
P. Larson, W. R. L. Lambrecht, A. Chantis, M. van Schilfgaarde
Phys. Rev. B {\bf 75} (2007) 045114.
%
%
\bibitem{rf:GdN3}
A. R. H. Preston, B. J. Ruck, W. R. L. Lambrecht, L. F. J. Piper, J. E. Downes, K. E. Smith, H. J. Trodahl,
App. Phys. Lett. {\bf 96} (2010) 032101.
%
%
\bibitem{rf:GdN4}
H. M. Liu, C. Y. Ma, C. Zhu, J-M. Liu,
J. Phys.: Condens. Matter {\bf 23} (2011) 245901.
%
%
\bibitem{rf:YIG1}
W. Y. Ching, Z-Q Gu, and Y-N. Xu,
J. App. Phys. {\bf 89} (2001) 6883.
%
%
\bibitem{rf:YIG2}
X. Jia, K. Liu, K. Xia, G. E. W. Bauer,
arXiv:1103.3764 (2011).
%
%
\bibitem{rf:Beenakker}
C. W. J. Beenakker,
Phys. Rev. Lett.  {\bf 67} (1991) 3836.
%
%
%
%
%
%
%
%
%
%
\end{thebibliography}
\end{document}